\documentclass[preprint,pra,aps,showkeys]{revtex4}
\usepackage{amsmath,amssymb}
\usepackage{graphicx}
\usepackage{color}
\usepackage{dcolumn}
\usepackage{bm}
\usepackage{hyperref}
\hypersetup{colorlinks=true,linkcolor=blue,citecolor=red}
\begin{document}

\title{Ultimate Precision of Direct Tomography of Wave Functions}

\author{Xuan-Hoai Thi Nguyen}
\author{Mahn-Soo Choi}
\email{choims@korea.ac.kr}
\affiliation{Department of Physics, Korea University, Seoul 02841, South Korea}

\date{\today}

\begin{abstract}
In contrast to the standard quantum state tomography, the direct tomography
seeks the direct access to the complex values of the wave function at
particular positions (i.e., the expansion coefficient in a fixed
basis). Originally put forward as a special case of weak measurement, it can
be extended to arbitrary measurement setup. We generalize the idea of
``quantum metrology,'' where a real-valued phase is estimated, to the
estimation of complex-valued phase, and apply it for the direct tomography of
the wave function. It turns out that the reformulation can help us easily find
the optimal measurements for efficient estimation.
We further propose two different measurement schemes that
eventually approach the Heisenberg limit. In the first scheme, the ensemble of
measured system is duplicated and the replica ensemble is time-reversal
transformed before the start of the measurement. In the other method, the
pointers are prepared in special entangled states, either GHZ-like maximally
entangled state or the symmetric Dicke state. In both methods, the real part
of the parameter is estimated with a Ramsey-type interferometry while the
imaginary part is estimated by amplitude measurements. The optimal condition
for the ultimate precision is achieved at small values of the complex
parameters, which provides possible explanations why the previous
weak-measurement scheme was successful.
\end{abstract}


\maketitle

\newpage

\newcommand\bra[1]{\mathinner{\langle{\textstyle#1}\rvert}}
\newcommand\ket[1]{\mathinner{\lvert{\textstyle#1}\rangle}}
\newcommand\braket[1]{\mathinner{\langle{\textstyle#1}\rangle}}
\newcommand\avg[1]{\mathinner{\langle{\textstyle#1}\rangle}}
\newcommand\hatA{\hat{A}}
\newcommand\hatI{\hat{I}}
\newcommand\hatJ{\hat{J}}
\newcommand\hatK{\hat{K}}
\newcommand\hatM{\hat{M}}
\newcommand\half{\frac{1}{2}}
\newcommand\rme{\mathrm{e}}
\newcommand\rmi{\mathrm{i}}
\newcommand\im{\operatorname{Im}}
\newcommand\re{\operatorname{Re}}
\newcommand\red{\color{red}}
\newcommand\mscnote[1]{\textcolor{red}{MSC: #1}}

\section{Introduction}
\label{Paper::sec:1}

Reconstruction of the quantum state of a system is of vital importance not
only in fundamental studies of quantum mechanics but also in many practical
application of quantum information technology. The standard way to do it, the
so-called quantum state tomography, requires an indirect computational
reconstruction based on the measurement outcomes of a complete set of
non-commuting observables on identically prepared
systems~\cite{Lvovsky09a}. Recently, an alternative method to has been put
forward and demonstrated experimentally~\cite{Lundeen11a,Lundeen12a}.  It has
attracted much interest because it enables the complex-valued wave functions
to be extracted directly and, from many points of view, in an experimentally
less challenging manner.  We call this method as the \emph{direct tomography}
of wave functions.
The direct tomography was originally proposed as a special case of weak
measurement in which the system is weakly coupled with the pointer,
post-selected on to a fixed state, and finally the wave functions are directly
access from the joint probabilities of some projective measurements on the
pointer~\cite{Lundeen11a,Lundeen12a}. Later it was extended to arbitrary
measurement setup working regardless of the system-pointer coupling strength
\cite{Vallone16a,Calderaro18a}. More recently, the direct tomography has been
reinterpreted in the so-called probe-controlled system framework. The latter
allows experimenters for even wider variations of setup and in many cases
leads to bigger efficiency such a scan-free method of direct
tomography~\cite{Ogawa19a}. However, the metrological aspect of the direct
tomography has not been paid attention yet.

The statistical nature sets the standard quantum limit on the precision of
standard measurement techniques~\cite{Braunstein94a,Braunstein96a}. To reduce
the statistical error, one needs to perform a large number $N$ of repeated
measurements. When it comes to direct tomography, it seems even worse as the
post-selection procedure demands even more repetition of experiments.
Recent efforts in quantum metrology have shown new insights to overcome the
standard quantum limit and achieve higher precision measurements by exploiting
quantum resources, especially, quantum
entanglement~\cite{Braunstein94a,Braunstein96a,Giovannetti04a,Giovannetti06a}.
A great number of measurement strategies along the line have been proposed and
demonstrated experimentally so far~\cite[see, e.g.,][and references
therein]{Pezze18a}. It has been found that there is a connection between
entanglement and high precision metrology\cite{Pezze09a}. Notably, the genuine
multi-particle entanglement is needed to achieve the maximum precision, the
so-called Heisenberg limit\cite{Giovannetti06a}.

In this paper, we investigate the ultimate precision of the direct tomography
of wave functions. For the purpose, we generalize the idea of quantum
metrology to the estimation of complex-valued phase, and apply it for the
direct tomography; see Section~\ref{Paper::sec:2}. We show that the
reformulation enables to easily find the optimal measurements for efficient
estimation.
We further propose two different measurement schemes that eventually approach
the Heisenberg limit (Section~\ref{Paper::sec:3}). In the first method, the
pointers are prepared in special entangled states, either GHZ-like maximally
entangled state (Section~\ref{Paper::sec:3.1}) or the symmetric Dicke state
(Section~\ref{Paper::sec:3.2}). In the other scheme, the ensemble of measured
system is duplicated and the replica ensemble is time-reversal transformed
before the start of the measurement (Section~\ref{Paper::sec:3.3}). In both
methods, the real part of the parameter is estimated with a Ramsey-type
interferometry while the imaginary part is estimated by amplitude
measurements.

\section{Direct tomography as a phase estimation}
\label{Paper::sec:2}

In order to investigate the precision limit of the direct tomography of wave
functions, it is convenient to reformulate it as a phase estimation in quantum
metrology. It allows clearer picture of the optimal initial states and
measurements on the pointers.

Before the reformulation, we briefly summarize the procedure of the direct tomography \cite{Lundeen11a,Lundeen12a,Vallone16a}. Note that here we follow Ref.~\cite{Vallone16a} and examine the direct tomography beyond weak-coupling approximation. 
Consider an unknown pure state $\ket{\psi_\text{S}}$ in a $d$
dimensional Hilbert space and expand it in a given basis
$\{\ket{x}|x=1,\cdots,d\}$ as
\begin{equation}
\label{1}
\ket{\psi_\mathrm{S}}=\sum_{x=1}^d\psi_x\ket{x}.
\end{equation}
A qubit is taken as the pointer and prepared in the state
$\ket{\phi_\mathrm{in}}$.
The total wave function of the system plus the pointer is thus $\ket{\Psi_\mathrm{in}}=\ket{\psi_S}\otimes\ket{\phi_\mathrm{in}}$.
The direct measurement of wave function $\psi_x$ starts by coupling the system with the pointer. The system-pointer interaction can be described by a unitary operator of the form
\begin{equation}
\label{Paper::eq:1}
\hat U_x
=\rme^{-\rmi\theta\ket{x}\bra{x}\otimes\hatK/2}
=(\hat{I}_\text{S}-\ket{x}\bra{x})\otimes\hat{I}_\text{P}+\ket{x}\bra{x}\otimes\rme^{-\rmi\theta\hatK/2},
\end{equation}
where $\theta$ is the system-probe coupling constant, $\hatK/2$ is a traceless
``angular momentum'' operator (i.e., $e^{-i\theta\hatK/2}$ is an ``rotation''
operator) on the probe, and $\hat{I}_\text{S}$ ($\hat{I}_\text{P}$) is the
identity operator on the system (probe). After the interaction, the system is
post-selected on to the state
\begin{equation}
\label{Paper::eq:3}
|p_0\rangle=\frac{1}{\sqrt{d}}\sum_{x=1}^d\ket{x} ,
\end{equation} 
leaving the pointer in the state
\begin{equation}
\label{Paper::eq:2}
\ket{\phi_\mathrm{f}} = \frac{1}{\sqrt{|\alpha|^2+|\beta|^2}}
\left(\alpha\hatI_P - \rmi\beta\hatK\right)\ket{\phi_\mathrm{in}} \,,
\end{equation}
where
\begin{equation}
\label{Paper::eq:4}
\alpha=\frac{\tilde{\psi}-\psi_x+\psi_x\cos(\theta/2)}{\sqrt{d}} \,,\quad
\beta=\frac{\psi_x\sin(\theta/2)}{\sqrt{d}} \,,\quad
\tilde{\psi}=\sum_x\psi_x.
\end{equation}
We assume $\tilde{\psi}\neq 0$ without loss
of generality; if $\tilde{\psi}= 0$, then one can choose a post-selection to a
different state. Moreover, the global phase can be arbitrarily chosen so that
$\tilde{\psi}$ is real valued and positive.
To extract the complex-valued wave function $\psi_x$, one measures three
observables: One is $\hatK/2$ in the system-probe coupling and the other two
are another angular momentum operators, $\hatK_1/2$ and $\hatK_2/2$,
perpendicular to $\hatK$. Through a number of independent measurements, the
probabilities $P_M$ of the measurement outcome $1$ (contrary to $-1$) for
measurement $M=K,K_1,K_2$ are inferred and then the wave function is given by
\begin{equation}
\label{Paper::eq:5}
\psi_x = \frac{d}{\tilde\psi\sin\theta}\left[
\left((1-P_{K_1})\tan\frac{\theta}{2}+P_{K_2}-\half\right)
+ \rmi\left(P_K-\half\right)
\right].
\end{equation}
Therefore, in principle, the wave function $\psi_x$ is estimated exactly as
long as the probabilities $P_M$ are inferred out of an infinite number of
repeated measurements. In practice, however, the number of repeated measurements are finite and the accuracy is subject to the standard quantum limit.

Now let us reformulate the direct tomography outlined above as a phase
estimation problem. To this end, we rewrite the normalized pointer state
after post-selection into the form
\begin{equation}
\label{phif}
\ket{\phi_\mathrm{f}}
=\sqrt\frac{\alpha^2+\beta^2}{|\alpha^2+\beta^2|}\frac{\rme^{-\rmi\varphi\hatK/2}|\phi_\mathrm{in}\rangle}{\sqrt{\langle\phi_\mathrm{in}|\rme^{\rmi(\varphi^*-\varphi)\hatK/2}|\phi_\mathrm{in}\rangle}}.
\end{equation}
where we have introduced a complex-valued phase
$\varphi$ by the relations
\begin{equation}
\cos\frac{\varphi}{2}=\frac{\alpha}{\sqrt{\alpha^2+\beta^2}} \,,\quad
\sin\frac{\varphi}{2}=\frac{\beta}{\sqrt{\alpha^2+\beta^2}} \,.
\end{equation}
Once the complex-valued parameter $\varphi$ is estimated through experiments,
one can get the wave function $\psi_x$ in a straight forward manner
\begin{equation}
\psi_x=\frac{\tilde{\psi} \tan(\varphi/2)}{2\sin(\theta/4)[\cos(\theta/4)+\sin(\theta/4)\tan(\varphi/2)]}.
\end{equation}

Whereas the relation~\eqref{phif} between the final and initial state is
\emph{formally} the same as the standard phase estimation in quantum
metrology \cite{Giovannetti06a}, it involves two \emph{two} parameters,
$\varphi_1:=\re\varphi$ and $\varphi_2:=\im\varphi$, and corresponds to a
multi-parameter quantum metrology \cite{Szczykulska16a}.  Naturally, it
requires measurements of more than one observables. Throughout this work, the
estimation of complex parameter $\varphi=\varphi_1+\rmi\varphi_2$ will
be used interchangeably with the multi-parameter estimation of real parameters
$\varphi_1$ and $\varphi_2$.

To see how to estimate the complex-valued phase $\varphi$ in
Eq.~\eqref{phif}, we note that
\begin{equation}
\label{Re}
\begin{bmatrix}
\langle\hatK_1\rangle_\mathrm{f}\\
\langle\hatK_2\rangle_\mathrm{f}
\end{bmatrix}
=
\frac{1}{\cosh(\varphi_2)+\sinh(\varphi_2)\avg{\hatK}_\mathrm{in}}
\begin{bmatrix}
\cos(\varphi_1) & -\sin(\varphi_1)\\
\sin(\varphi_1) &  \cos(\varphi_1)
\end{bmatrix}
\begin{bmatrix}
\langle\hatK_1\rangle_\mathrm{in}\\
\langle\hatK_2\rangle_\mathrm{in}
\end{bmatrix}
\end{equation}
and
\begin{equation}
\label{Im}
\avg{\hatK}_\mathrm{f}
=\frac{ \sinh(\varphi_2) +
  \cosh(\varphi_2)\avg{\hatK}_\mathrm{in} }{
  \cosh(\varphi_2)+\sinh(\varphi_2)\avg{\hatK}_\mathrm{in}},
\end{equation}
where $\langle...\rangle_\mathrm{f}$ denotes the statistical average $\langle\phi_\mathrm{f}|...\ket{\phi_\mathrm{f}}$ and analogously $\langle...\rangle_\mathrm{in}$.
It is observed from Eq.~\eqref{Re} and \eqref{Im} that
$\operatorname{Re}(\varphi)$ rotates the classical vector
$(\langle\hatK_1\rangle,\langle\hatK_2\rangle)$ around the axis along $\hatK$
whereas $\operatorname{Im}(\varphi)$ shifts $\avg{\hatK}$. Such a
rotation angle $\operatorname{Re}(\varphi)$ can be estimated by a Ramsey-type
interferometry whereas the estimation of $\operatorname{Im}(\varphi)$ requires
an amplitude measurement scheme.
In particular, for a choice of $\ket{\phi_\mathrm{in}}$ consistent with the
optimal sensitivity such that
$\avg{\hatK_2}_\mathrm{in}=\avg{\hatK}_\mathrm{in}=0$, one has
\begin{equation}\label{Imb}
\avg{\hatK_1}_\mathrm{f} = \frac{\cos(\varphi_1)}{\cosh(\varphi_2)}
\avg{\hatK_1}_\mathrm{in} \,,\quad
\avg{\hatK}_\mathrm{f} = \tanh(\varphi_2).
\end{equation}
In short, the optimal estimation of the complex-valued phase $\varphi$,
one needs first to (i) prepare the probe in the initial state such that
$\avg{\hatK_2}=\avg{\hatK}=0$, and then (ii) perform measurements of
\emph{two} [rather than three as in Eq.~\eqref{Paper::eq:5}] observables $\hatK_1$ and $\hatK$. Therefore, the reformulation of direct tomography in the form of complex phase estimation is intuitively appealing and helps us find the optimal measurements for efficient estimation.

Below we show that by preparing a multi-qubit probe in entangled states one
can achieve the Heisenberg limit for the complex-valued phase estimation (and
hence the complex-valued wave functions).

\section{Precision limits of the direct tomography}
\label{Paper::sec:3}

It is known that the estimation of a real-valued phase can reach the Heisenberg limit by exploiting quantum entanglements in the pointers \cite{Giovannetti06a}. The question is whether the same limit can be achieved for the estimation of a complex-valued phase involved in the direct tomography. Here we demonstrate that a complex-valued phase can be estimated in the Heisenberg limit by preparing multi-qubit pointers in quantum entanglement and choosing proper measurements.

\subsection{Using N00N state}
\label{Paper::sec:3.1}

Consider an ensemble of $N$ systems all in the same state $\ket{\psi_S}$.  We
take a set of $N$ qubits as the pointers and prepare them in the so-called
NOON state (or the $N$-qubit GHZ state),
\begin{equation}
\ket{\phi_\mathrm{in}} =
\frac{|0\rangle^{\otimes N}+|1\rangle^{\otimes N}}{\sqrt{2}} ,
\end{equation}
which has proved particularly
interesting in high-precision quantum metrology \cite{Pezze18a}.
We couple each system in the ensemble to each corresponding pointer qubit so that the overall unitary operator of the interaction is given by
\begin{equation}
\hat{U}_x^{\otimes N}=\left[\exp\left(-\rmi\theta\ket{x}\bra{x}\otimes\frac{\hat\sigma_z}{2}\right)\right]^{\otimes N}.
\end{equation}
Here we have chosen $\hatK=\hat\sigma_z$ to be concrete.  After
post-selecting every system on to the state $\ket{p_0}$ in
Eq.~\eqref{Paper::eq:3}, the (normalized) final state of the pointers is given
by
\begin{equation}
\label{NP}
\ket{\phi_\mathrm{f}}_\text{N00N}=\frac{(\alpha-\rmi\beta)^N|0\rangle^{\otimes N}+(\alpha+\rmi\beta)^N|1\rangle^{\otimes N}}{\sqrt{|\alpha-\rmi\beta|^{2N}+|\alpha+\rmi\beta|^{2N}}}
\end{equation}
with $\alpha$ and $\beta$ defined in Eq.~\eqref{Paper::eq:4}. Equivalently, in
accordance with the phase-estimation formulation~\eqref{phif}, it can be
rewritten as $\ket{\phi_\mathrm{f}}_\text{N00N}$ as follows (up to a global
phase):
\begin{equation}
\ket{\phi_\mathrm{f}}_\text{N00N}
=\frac{\rme^{-\rmi N\varphi/2}|0\rangle^{\otimes N}
  +\rme^{\rmi N\varphi/2}|1\rangle^{\otimes N}}{\sqrt{2 \cosh(N\im\varphi)}}.
\end{equation}

Now consider two measurements $\hatM_1:=\hat{\sigma}_x^{\otimes N}$ and
$\hatM_2:=\hat{\sigma}_z^{\otimes N}$. We note that
\begin{align}
\langle \hatM_1\rangle:=\langle\hat{\sigma}_x^{\otimes N}\rangle
&= \frac{\cos(N\re\varphi)}{\cosh(N\im\varphi)},\\
\langle \hatM_2\rangle:=\langle\hat{\sigma}_z^{\otimes N}\rangle
&= \tanh(N\im\varphi).
\end{align}
Assuming small variations of the measurements with the parameter $\varphi=\varphi_1+\rmi\varphi_2$,
the covariance matrix $C_{ij}(\varphi):=\Delta\varphi_i\Delta\varphi_j$
($i,j=1,2$) of the estimators $\varphi_1=\re\varphi$ and
$\varphi_2=\im\varphi$ is related to the covariance of the measurement
$\avg{\Delta \hatM_\mu\Delta \hatM_\nu}$ ($\mu,\nu=1,2$) by the error-propagation formula
\begin{equation}
\label{EP}
\avg{\Delta \hatM_\mu\Delta \hatM_\nu}
= \sum_{ij}
\frac{\partial\langle \hatM_\mu\rangle}{\partial\varphi_i}
C_{ij}(\varphi)
\frac{\partial\langle \hatM_\nu\rangle}{\partial\varphi_j}.
\end{equation}
Inverting the error propagation formula, we find that the precision is given
by
\begin{equation}
\label{V1}
(\Delta\varphi_1)^2=(\Delta\varphi_2)^2=\frac{\cosh^2(N\varphi_2)}{N^2}.
\end{equation}
It is concluded that $\hatM_1$ and $\hatM_2$ are indeed optimal measurements
under the optimal condition $N\varphi_2\to 0$ for the Heisenberg limit.  Here
we have chosen specific measurements $\hatM_1$ and $\hatM_2$, but more general
argument in terms of the Fisher information and the Cramer-Rao bound; see
Appendix~\ref{Paper::sec:A}.

\subsection{Using Dicke state}
\label{Paper::sec:3.2}

Thanks to their experimental relevance, the symmetric Dicke states have also
been widely used for quantum entanglement~\cite{Pezze18a}.  In particular, it
was illustrated that the entanglement in a Dicke state enables one to achieve
the Heisenberg-limited interferometry for the single-parameter quantum
metrology~\cite{Apellaniz15a}.  Given $N$ qubits, the
symmetric Dicke state $\ket{j\equiv N/2,m}$ with $m=j,j-1,\cdots,-j$ is defined by
\begin{equation}
\ket{j,m}
:=\sqrt\frac{(j-m)!}{(2j)!}\sum_P
\hat{P}|\underbrace{11...}_{j-m}\underbrace{00...}_{j+m}\rangle,
\end{equation}
where the sum is over all possible permutations $P$ and $\hat{P}$
is the corresponding permutation operator.

We proceed in a similar manner as with the initial NOON state of pointers.
The pointers of $N$ qubits are initially prepared in the particular Dicke state
$\ket{\phi_\mathrm{in}}=\ket{j\equiv N/2,0}$. Each pointer is coupled with a system in the ensemble so that the unitary interaction is given by
\begin{equation}
\hat{U}_x^{\otimes N}
=\left[\exp(-\rmi\theta\ket{x}\bra{x}\otimes\hat\sigma_y/2)\right]^{\otimes N}.
\end{equation}
Here we have chosen $\hatK=\hat\sigma_y$ to make the best use of the
characteristic of the Dicke state; namely, the sharp distribution along the
equator of the generalized Bloch sphere \cite{Pezze18a}.  After
post-selection, the final state \eqref{phif} of the pointers becomes
\begin{equation}\label{Dicke}
\ket{\phi_\mathrm{f}}_\text{Dicke}
=\frac{\rme^{-\rmi\varphi\hat{J}_y}|j,0\rangle}{\sqrt{\langle j,0|\rme^{\rmi(\varphi^*-\varphi)\hat{J}_y}|j,0\rangle}},
\end{equation}
where 
\begin{equation}
\hat J_\mu=\frac{1}{2}\sum^N_{k=1}\hat{\sigma}^{(k)}_\mu
\quad (\mu=x,y,z) .
\end{equation}
For later use, we define the Wigner matrix element
\begin{equation}
W_{mm'}^{(j)} := \bra{j,m}e^{-i\varphi\hatJ_y}\ket{j,m'}
\end{equation}
Here note that the phase $\varphi$ is complex in general.
For integer $j$, the expression for the matrix element $W_{m0}^{(j)}$ is especially simple as
\begin{equation}\label{WM}
W^{(j)}_{m0}(\varphi) = P^m_j(\cos\varphi)\sqrt{\frac{(j-m)!}{(j+m)!}}
\quad (\re\varphi>0) \,,
\end{equation}
where $P^m_j(z)$ denotes the associated Legendre polynomial of argument $z$.

Unlike the NOON state, the Dicke state does not allow for simple expressions
for the Fisher information and the corresponding Cramer-Rao bound. Instead, we
choose the optimal measurements based on the characteristics of the Dicke
state and its behavior under the collective rotation $e^{-i\varphi\hatJ_y}$ by
a complex angle $\varphi$. As mentioned above, the Dicke state has a sharp
distribution along the equator of the Bloch sphere. Then
$e^{-i\re\varphi\hatJ_y}$ rotates this distribution off the equator. The
resulting sharp contrast with the initial state state can be detected most
efficiently by measuring $\hatJ_z^2$. On the other hand,
$e^{\im\varphi\hatJ_y}$ tends to pull the distribution along the positive
$y$-axis. This deviation can be efficiently detected by measuring
$\hatJ_y$. Below we demonstrate that $\hatJ_y$ and $\hatJ_z^2$ are indeed optimal measurements to achieve the Heisenberg limit.

We start with the analysis of the measurement of $\hatJ_y$: By virtue of the theory of angular momentum, we acquire
\begin{align}
\label{Jy}
\langle\hat{J}_y\rangle
&= \frac{\rmi W^{(j)}_{10}(2\rmi\varphi_2)}{W^{(j)}_{00}(2\rmi\varphi_2)}\sqrt{j(j+1)}, \\
\label{Jy2}
\langle\hat{J}^2_y\rangle
&= j(j+1) -
\frac{\rmi W^{(j)}_{10}(2\rmi\varphi_2)}{W^{(j)}_{00}(2\rmi\varphi_2)}
\coth(2\varphi_2)\sqrt{j(j+1)}.
\end{align}
It then follows from the error-propagation formula that
\begin{align}
\frac{1}{(\Delta\varphi_2)^2}
&= \frac{1}{(\Delta\hat{J}_y)^2}
\left(\frac{\partial\avg{\hat{J}_y}}{\partial\varphi_2}\right)^2
=4(\Delta\hat{J}_y)^2\nonumber\\
&= 4j(j+1)\left[1 -
\frac{\rmi W^{(j)}_{10}(2\rmi\varphi_2)}{W^{(j)}_{00}(2\rmi\varphi_2)}
\frac{\coth(2\varphi_2)}{\sqrt{j(j+1)}} -
\left(\frac{\rmi W^{(j)}_{10}(2\rmi\varphi_2)}
{W^{(j)}_{00}(2\rmi\varphi_2)}\right)^2\right].
\label{er}
\end{align}
Equation~\eqref{er} implies that the larger $(\Delta\hat{J}_y)^2$ is the more
precise the estimation of $\im\varphi$ gets, which leads to the optimal
condition $\im\varphi=0$.
Putting the optimal condition into Eq.~\eqref{er} gives the Heisenberg limit
\begin{equation}
\label{Hsb}
(\Delta\im\varphi)^2_{\mathrm{opt}}=\frac{2}{N(N+2)}
\end{equation}
for the estimation of $\im\varphi$.  It is interesting to note that the
variance $(\Delta\varphi_2)^2$ in Eq.~\eqref{er} depends only on $\varphi_2$
but not on $\varphi_1$. This is another important feature that allows
$\varphi_2$ to be estimated independently of $\varphi_1$ through the
measurement $\hatJ_y$.

\begin{figure}[tbp]
\centering
\includegraphics[width=0.65\linewidth]{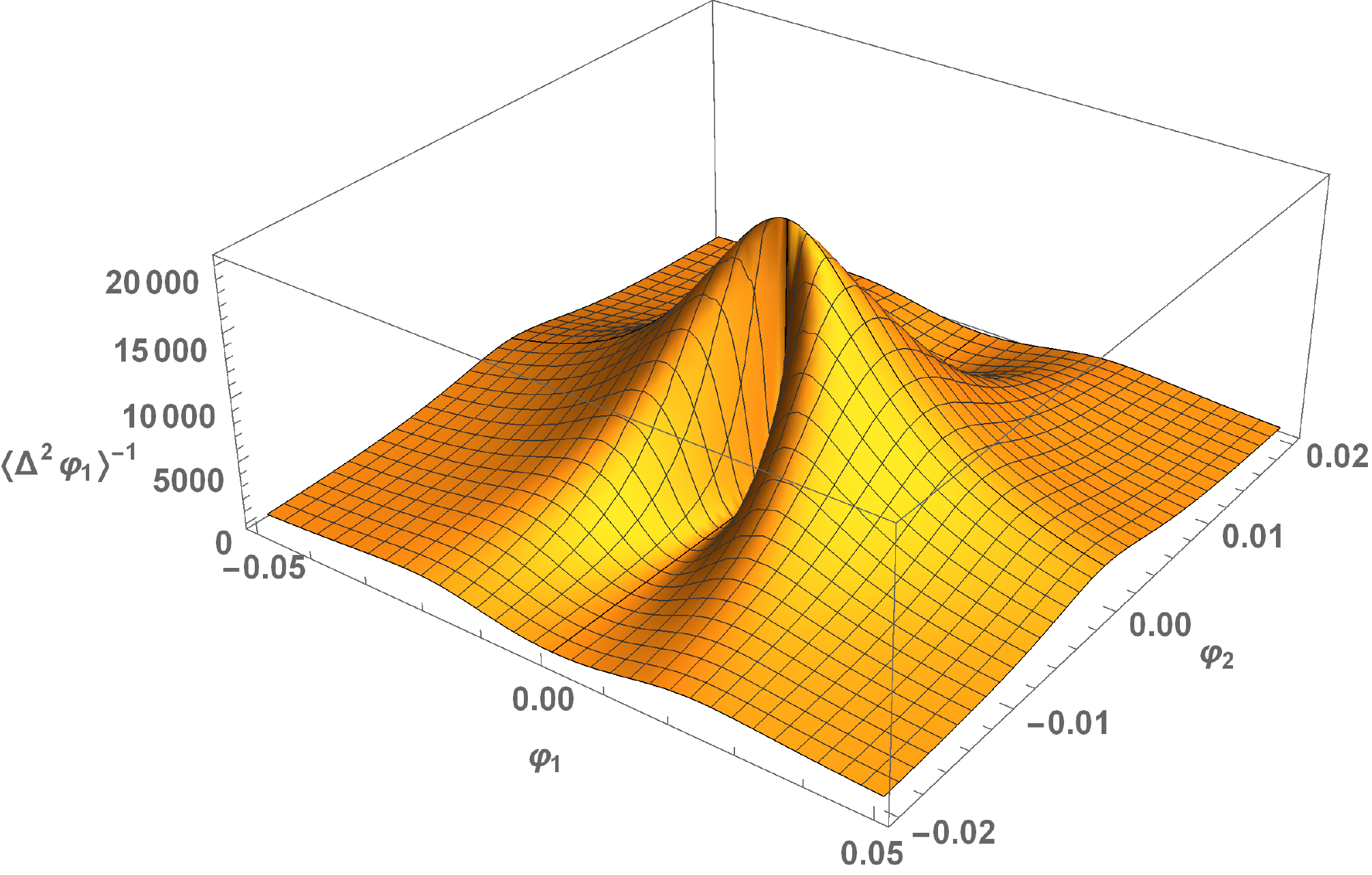}
\includegraphics[width=0.60\linewidth]{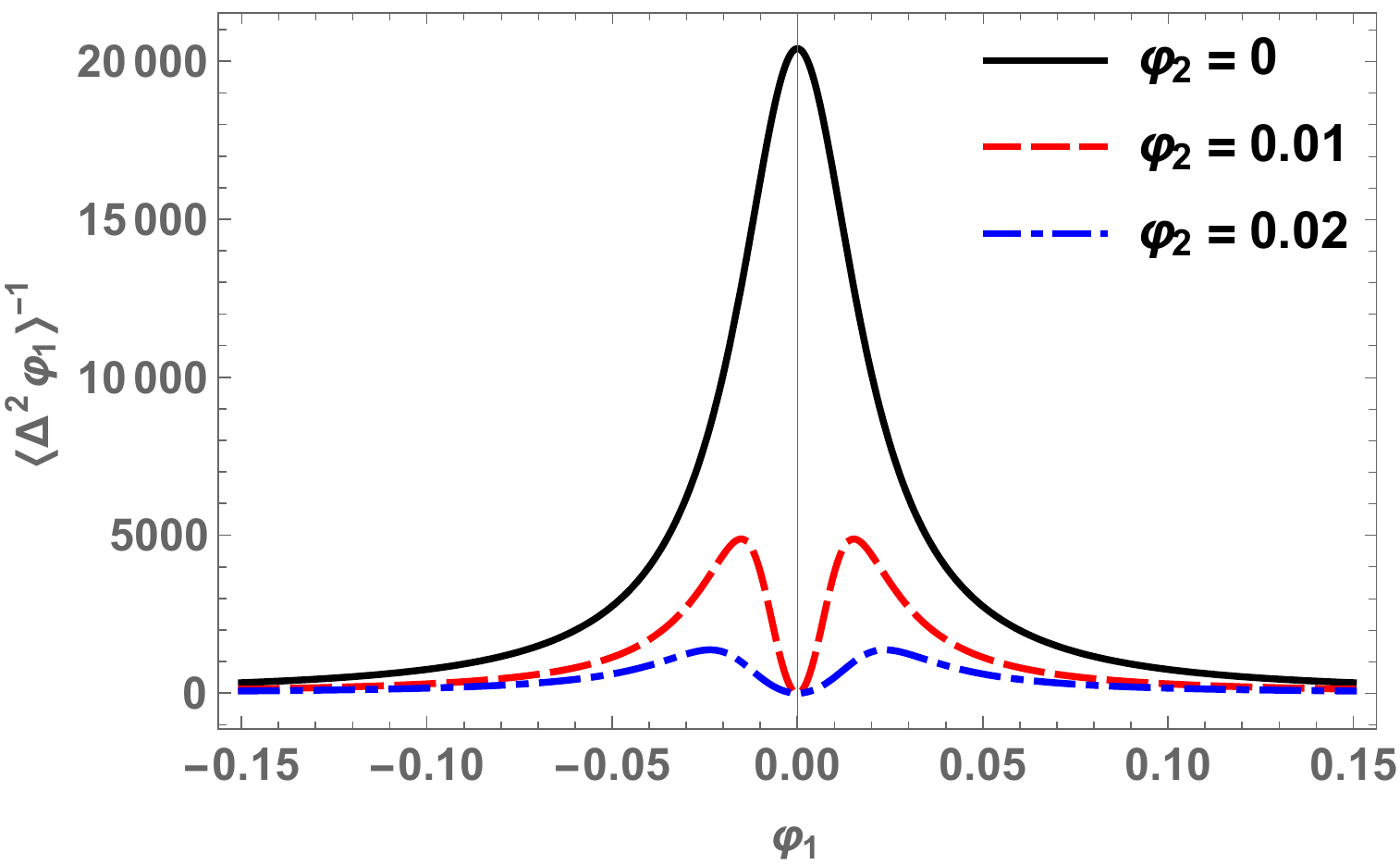}
\caption{(a) The three-dimensional plot of the reciprocal of
  $(\Delta\re\varphi)^2$ for the points of $N=50$ qubits in the Dicke
  state. (b) The same plot as a function of $\re\varphi$ at several fixed
  values of $\im\varphi$.}
\label{fig1}
\end{figure}

To analyze the measurement $\hatJ_z^2$ as an estimator of $\varphi_1$, we
evaluate
\begin{align}
\label{Jz}
\avg{\hat{J}^2_z}
&= \frac{\rmi W^{(j)}_{10}(2\rmi\varphi_2)}{W^{(j)}_{00}(2\rmi\varphi_2)}
\sqrt{j(j+1)}\sin(\varphi)
\left[\coth(2\varphi_2)\sin(\varphi)-\rmi\cos(\varphi)\right] \\
\label{Paper::eq:6}
\langle \hatJ^4_z\rangle
&= \frac{1}{W^{(j)}_{00}(2\rmi\varphi_2)}
\sum_{m=-j}^j m^4| W^{(j)}_{m0}(\varphi)|^2 .
\end{align}
Unlike the fortunate case with the measurement $\hatJ_y$, the moments
$\avg{\hatJ_z^2}$ and $\avg{\hatJ_z^4}$ depend not only on $\varphi_1$ but
also on $\varphi_1$. Therefore one has to use the multi-parameter
error-propagation formula~\eqref{EP} with $\hatM_1=\hatJ_z^2$ and
$\hatM_2=\hatJ_y$, which leads to
\begin{equation}
\label{Var1}
(\Delta\varphi_1)^2
=\frac{\avg{(\Delta\hatJ_z^2)^2}^2+\left(\frac{\partial\langle
    \hatJ^2_z\rangle}{\partial\varphi_2}\right)^2(\Delta\varphi_2)^2
-\frac{\partial\langle\hatJ^2_z\rangle}{\partial\varphi_2}\left(\frac{\partial\langle\hatJ_y\rangle}{\partial\varphi_2}\right)^{-1}
\avg{\{\Delta\hatJ^2_z,\Delta\hatJ_y\}}}{\left(\frac{\partial\langle \hatJ^2_z\rangle}{\partial\varphi_1}\right)^2},
\end{equation}
where we have defined $\Delta\hatA=\hatA-\avg{\hatA}$ for operator $\hatA$ and
noted that $\partial\langle\hatJ_y\rangle/\partial\varphi_1=0$.
We refer the technical details of its calculations to
Appendix~\ref{Paper::sec:B}, and instead summarize its behavior in
Fig.~\ref{fig1} as a function of $\varphi_1$ and $\varphi_2$ for the pointers
of $N=50$ qubits.
It is clear from Fig.~\ref{fig1} that the optimal condition is given by
$\varphi_1=\varphi_2=0$. At these optimal condition, the precision of
$\varphi_1$ is given by the Heisenberg limit
\begin{equation}
\label{Paper::eq:7}
(\Delta\varphi_1)^2_\mathrm{opt} = \frac{2}{N(N+2)}.
\end{equation}
Incidentally, by putting the optimal condition $\varphi_2=0$ obtained
independently through the measurement $\hatJ_y$ above, we get
\begin{equation}
\label{p1}
\left.(\Delta\varphi_1)^{-2}\right|_{\varphi_2=0}
= \frac{8j(j+1)}{(j^2+j-2)\tan^2(\varphi_1)+4},
\end{equation}
which coincides with the single-parameter estimation in
Ref.~\cite{Apellaniz15a} as it should.

\subsection{Using time-reversal ensemble}
\label{Paper::sec:3.3}

The reformulation of the direct tomography as an complex-valued phase estimation in~\eqref{phif} inspires another interesting strategy based on time-reversal (TR) transformation.
Given an ensemble of system in the state \eqref{1}, we prepare another ensemble in the TR state
\begin{equation}
|\bar{\psi}_\text{S}\rangle
:= \hat{T}\ket{\psi_\mathrm{S}}
= \sum_{x=1}^d\psi^*_x\ket{x},
\end{equation}
where $\hat{T}$ is the (anti-unitary) TR operator (here we assume for
simplicity that the basis state $\ket{x}$ is invariant under the TR
transformation).
The pointers of $2N$ qubits are prepared, say, in the NOON state.  The first
$N$ qubits interact with the systems in the original ensemble whereas the
other $N$ qubits are coupled with ones in the time-reversal ensemble.
After post-selection, the pointers get in the final state of the form
\begin{equation}
\label{TRS}
\ket{\phi_\mathrm{f}}_\text{TRS}
=\frac{[(\alpha-\rmi\beta)(\alpha^*-\rmi\beta^*)]^N|0\rangle^{\otimes 2N}+[(\alpha^*+\rmi\beta^*)(\alpha+\rmi\beta)]^N|1\rangle^{\otimes 2N}}{\sqrt{|(\alpha-\rmi\beta)(\alpha^*-\rmi\beta^*)|^{2N}+|(\alpha^*+\rmi\beta^*)(\alpha+\rmi\beta)|^{2N}}} .
\end{equation}
Now recall that for any complex variables $\alpha$ and $\beta$,
\begin{equation}
|(\alpha-\rmi\beta)(\alpha^*-\rmi\beta^*)|
= |(\alpha^*+\rmi\beta^*)(\alpha+\rmi\beta)|.
\end{equation}
It recasts Eq.~\eqref{TRS} to the quantum metrologically appealing form
\begin{equation}
\label{fin}
\ket{\phi_\mathrm{f}}_\text{TRS}
=\frac{|0\rangle^{\otimes 2N}
  + \rme^{\rmi2N\varphi_1}|1\rangle^{\otimes 2N}}{\sqrt{2}}.
\end{equation}
Namely, the above state is identical to the state with the amplified phase
shift in interferometries with the NOON state, one of the earliest
experimental demonstrations of the Heisenberg limit~\cite{Bollinger96a}.
It is also worth noting that unlike the above two schemes, in which the
estimation of $\varphi_1$ strictly depends on that of $\varphi_2$, the scheme
using the TR ensemble enables the real part $\varphi_1$ to be estimated
independently. To estimate the imaginary part $\varphi_2$, we can apply the
measurement strategy proposed in Section~\ref{Paper::sec:3.1}.

As the TR transformation is anti-unitary, it cannot be implemented physically in isolated systems. However, it is achievable by embedding the system in a larger system. Therefore, as long as the setup permits additional capability of controlling the system, the TR ensemble provides an efficient strategy for direct precision measurement of wave functions.

\section{Conclusions}
\label{Paper::sec:4}

Generalizing the idea of quantum metrology of phase estimation, we have
reformulated the direct tomography of wave functions as the estimation of
complex-valued phase. It has turned out that the new formulation is
intuitively appealing and inspires the proper choices of optimal
measurements. We have further proposed two different measurement schemes that
eventually approach the Heisenberg limit. In the first method, the pointers
are prepared in special entangled states, either GHZ-like maximally entangled
state or the symmetric Dicke state. In the other scheme, the ensemble of
measured system is duplicated and the replica ensemble is time-reversal
transformed before the start of the measurement. In both methods, the real
part of the phase is estimated with a Ramsey-type interferometry while the
imaginary part is estimated by amplitude measurements. The optimal condition
for the ultimate precision is achieved at small values of the complex
phases, which provides possible explanations why the previous
weak-measurement scheme was successful.

\appendix
\section{The Cramer-Rao Bound for the NOON State}
\label{Paper::sec:A}

Here we analyze the precision limit of the complex-valued phase estimation
based on the multi-parameter estimation in terms of the Fisher information
matrix and the corresponding Cramer-Rao
bounds~\cite{Szczykulska16a,Holevo82a}.

We first briefly summarize the multi-parameter quantum
metrology~\cite{Szczykulska16a,Hradil04a}. Suppose that we want to estimate a
set of unknown parameters $\{X_\mu|\mu=1,\cdots,L\}$ through the measurements
of a positive-operator valued measure (POVM),
$\{\hat\Pi_j|j=1,2,...,L^\prime\}$.
The covariance matrix
\begin{math}
C_{\mu\nu}(\{X_\lambda\}) =
\Delta X_\mu\,\Delta X_\nu
\end{math} 
satisfies the following inequality
\cite{Szczykulska16a}
\begin{equation}
\label{CRB}
C(\{X_\mu\})\geq\mathcal{F}^{-1}(\{\hat\Pi_j\}), 
\end{equation}
where $\mathcal{F}(\{\Pi_j\})$ is the Fisher information matrix (FIM)
associated with the probability distribution $\{p_j(\{X_\mu\})\}$ for the
measurements $\{\Pi_j\}$. The entries of the Fisher information matrix are
defined by \cite{Hradil04a}
\begin{equation}
F_{\mu\nu}=\sum_{j=1}^{L^\prime}\frac{\partial_\mu p_j\partial_\nu p_j}{p_j}
\end{equation}
with $\partial_{\mu}$ denoting $\partial/\partial X_{\mu}$.
In the case of complex parameters $Z_\mu=X_\mu+\rmi Y_\mu$, one can keep the
complex structure in the covariance matrix and the Fisher information
matrix. In this case, one constructs the covariance matrix by replacing each
element by the $2\times 2$ block
\begin{equation}
C_{\mu\nu} =
\begin{bmatrix}
\Delta Z_\mu\,  \Delta Z_\nu^* &  \Delta Z_\mu\,  \Delta Z_\nu \\
\Delta Z_\mu^*\,\Delta Z_\nu^* &  \Delta Z_\mu^*\,\Delta Z_\nu
\end{bmatrix}
\end{equation}
Similarly, the Fisher information matrix is defined with respect to
two derivatives $\partial/\partial{Z_\mu^*}$ and $\partial/\partial{Z_\mu}$ for
each $Z_\mu$.

Now let us apply the multi-parameter Cramer-Rao bound~\eqref{CRB} in our
problem of estimating the wave function $\psi_x$ in \eqref{NP}. Calculating on
the final pointer state \eqref{NP} we obtain the probabilities of the POVM
elements as follows
\begin{equation}
p_j=\langle\hat\Pi_j\rangle_\text{f,N00N}=\frac{A_j|\alpha-\rmi\beta|^{2N}+B_j|\alpha+\rmi\beta|^{2N}+2\text{Re}\left[C_j(\alpha^*+\rmi\beta^*)^N(\alpha+\rmi\beta)^N\right]}{|\alpha-\rmi\beta|^{2N}+|\alpha+\rmi\beta|^{2N}},
\end{equation}
where
\begin{equation}
A_j=\langle00...|\hat\Pi_j|00...\rangle ,\quad
B_j=\langle11...|\hat\Pi_j|11...\rangle ,\quad
C_j=\langle00...|\hat\Pi_j|11...\rangle \quad
\end{equation}
and hence 
\begin{equation}
\partial_{\psi_x} p_j=N\frac{A_j-B_j-C_j(\gamma^*)^N+C_j^*(\gamma^*)^{-N}}{(|\gamma|^N+|\gamma|^{-N})^2}\partial_{\psi_x} \log(\gamma)
\end{equation}
with $\gamma=(\alpha-\rmi\beta)/(\alpha+\rmi\beta)$.  Assuming
$|\gamma|\geq 1$ without loss of generality, we see that, as
$N\rightarrow \infty$, $\partial_{\psi_x}p_j\rightarrow N|\gamma|^{-N}$. As a
result, for measurements such that $A_j=0$, we find
\begin{equation}\label{F}
\left(\mathcal{F}^{-1}\right)_{\mu\nu}\propto \frac{|\gamma|^N}{N^2}.
\end{equation}
Therefore, as $|\gamma|\to1$, which conforms the optimal condition for the estimation of $\psi_x$, the Heisenberg limit is saturated.

\section{Variance of the real part in the scheme using Dicke state}
\label{Paper::sec:B}

In this Appendix we provides the technical details involved in the calculation
of the moments $\avg{\hatJ_z^2}$ and $\avg{\hatJ_z^4}$ in Eqs.~\eqref{Jz} and
\eqref{Paper::eq:6}, respectively, which are required in~Eq.~\eqref{Var1}.

The terms $\langle\hatJ_y\rangle$,
$\partial\langle\hatJ_y\rangle/\partial\varphi_2$, and $(\Delta\varphi_2)^2$
are given by \eqref{Jy} and \eqref{er}. To calculate the remaining terms in
\eqref{Var1}, it is useful to recall the transformation rule
\begin{equation}\label{trans}
\hatJ_\mu(\varphi)=\rme^{\rmi\varphi\hatJ_y}\hatJ_\mu\rme^{-\rmi\varphi\hatJ_y}=\cos(\varphi)\hatJ_\mu+\rmi[\hatJ_y,\hatJ_\mu]\sin(\varphi).
\end{equation}

First, let us evaluate the average $\langle \hatJ^2_z\rangle$ and its
derivatives. By virtue of \eqref{trans}, one can obtain
\begin{eqnarray}
\langle \hatJ^2_z\rangle&=&\frac{1}{W^{(j)}_{00}(2\rmi\varphi_2)}\langle j,0|\rme^{\rmi\varphi^*\hatJ_y}\hatJ^2_z\rme^{-\rmi\varphi\hatJ_y}|j,0\rangle\nonumber\\
&=&\frac{1}{W^{(j)}_{00}(2\rmi\varphi_2)}\sum_{m=-j}^j\langle j,0|\rme^{\rmi2\varphi_2\hatJ_y}|j,m\rangle\langle j,m|\rme^{\rmi\varphi\hatJ_y}\hatJ^2_z\rme^{-\rmi\varphi\hatJ_y}|j,0\rangle\nonumber\\
&=&\sum_{m=-j}^j\frac{W^{(j)}_{0m}(2\rmi\varphi_2)}{W^{(j)}_{00}(2\rmi\varphi_2)}\left[\sin^2\varphi\langle j,m|\hatJ^2_x|j,0\rangle-\frac{\sin(2\varphi)}{2}\langle j,m|\hatJ_x\hatJ_z|j,0\rangle\right]\nonumber\\
&=& \frac{\rmi W^{(j)}_{10}(2\rmi\varphi_2)}{W^{(j)}_{00}(2\rmi\varphi_2)}
\sqrt{j(j+1)}\sin(\varphi)
\left[\coth(2\varphi_2)\sin(\varphi)-\rmi\cos(\varphi)\right]
\end{eqnarray}
Noting that
\begin{equation}
\left[\sqrt{\frac{(j+2)!}{(j-2)!}}\frac{W^{(j)}_{20}(2\rmi\varphi_2)}{W^{(j)}_{00}(2\rmi\varphi_2)}+j(j+1)\right]\sin(\rmi2\varphi_2)+2\sqrt{\frac{(j+1)!}{(j-1)!}}\frac{W^{(j)}_{10}(2\rmi\varphi_2)}{W^{(j)}_{00}(2\rmi\varphi_2)}\cos(\rmi2\varphi_2)=0,
\end{equation}
which is derived from the recurrence formula of the associated Legendre polynomial, $\langle \hatJ^2_z\rangle$ can be reduced to \eqref{Jz}. Taking the derivative of $\langle \hatJ^2_z\rangle$ given by \eqref{Jz} with respect to $\varphi_1$ and $\varphi_2$, respectively, we obtain
\begin{eqnarray}
\frac{\partial\langle \hatJ^2_z\rangle}{\partial\varphi_1}
&=&\frac{2\sin(2\varphi_1)}{\sinh(\varphi_2)}\frac{\rmi W^{(j)}_{20}(2\rmi\varphi_2)}{W^{(j)}_{00}(2\rmi\varphi_2)}\sqrt{j(j+1)},\\
\frac{\partial\langle \hatJ^2_z\rangle}{\partial\varphi_2}
&=&j(j+1)\left[\coth(2\varphi_2)-\frac{\cos(2\varphi_1)}{\sinh(2\varphi_2)}\right]\left[1-\left(\frac{\rmi W^{(j)}_{20}(2\rmi\varphi_2)}{W^{(j)}_{00}(2\rmi\varphi_2)}\right)^2\right]\nonumber\\
&&+\frac{\rmi W^{(j)}_{20}(2\rmi\varphi_2)}{W^{(j)}_{00}(2\rmi\varphi_2)}\sqrt{j(j+1)}\left[\frac{2\cos(2\varphi_1)\cosh(2\varphi_2)-2}{\sinh^2(2\varphi_2)}-1\right].
\end{eqnarray}

On the other hand, $\avg{\hatJ^4_z}$ can be expressed in terms of the Wigner matrix elements as following
\begin{align}
\langle \hatJ^4_z\rangle
&= \frac{1}{W^{(j)}_{00}(2\rmi\varphi_2)} \bra{j,0}
\rme^{\rmi\varphi^*\hatJ_y}\hatJ^4_z\rme^{-\rmi\varphi\hatJ_y}
\ket{j,0} \nonumber\\
&= \frac{1}{W^{(j)}_{00}(2\rmi\varphi_2)}
\sum_{m,m'}W^{(j)}_{m'0}(\varphi^*)W^{(j)}_{m0}(\varphi)\langle j,m'|\hatJ^4_z|j,m\rangle\nonumber\\
&= \frac{1}{W^{(j)}_{00}(2\rmi\varphi_2)}
\sum_{m=-j}^j m^4| W^{(j)}_{m0}(\varphi)|^2.
\end{align}

\bibliography{Paper}

\end{document}